\documentclass{pasj01}
\Received{$\langle$reception date$\rangle$}
\Accepted{$\langle$acception date$\rangle$}
\Published{$\langle$publication date$\rangle$}
\usepackage[switch,mathlines]{lineno}
\usepackage{ulem}

\begin{document}

\title{Intrinsic Alignments and Spin Correlations of [OII] Emitters at $z=1.2$ and $z=1.5$ from HSC Narrow-band Survey}

\author{Motonari \textsc{Tonegawa}\altaffilmark{1}, Teppei \textsc{Okumura}\altaffilmark{2,3} and Masao \textsc{Hayashi}\altaffilmark{4}}
\altaffiltext{1}{Asia Pacific Center for Theoretical Physics, Pohang, 37673, Korea}
\altaffiltext{2}{Institute of Astronomy and Astrophysics, Academia Sinica, No. 1, Section 4, Roosevelt Road, Taipei 106216, Taiwan}
\altaffiltext{3}{Kavli Institute for the Physics and Mathematics of the Universe (WPI), UTIAS, The University of Tokyo, 5-1-5 Kashiwanoha, Kashiwa, Chiba 277-8583, Japan}
\altaffiltext{4}{National Astronomical Observatory of Japan
2-21-1 Osawa, Mitaka, Tokyo 181-8588, Japan}
\email{motonari.tonegawa@apctp.org}


\maketitle

\begin{abstract}
Galaxies are known to be aligned toward speciﬁc directions within the large-scale structure.
Such alignment signals become important
for controlling the systematics of weak lensing surveys and for constraining galaxy formation and evolution scenarios.
We measure the galaxy-ellipticity and ellipticity-ellipticity correlation functions for blue star-forming galaxies at $z=1.19$ and $z=1.47$ that are selected by detecting [OII] emission lines in narrow-band filters of the Hyper Suprime-Cam on the Subaru Telescope. Assuming that disk galaxies are thin and rotation-supported, we also measure the spin correlation function by estimating spin directions with ellipticities and position angles. 
Above $1 \; h^{-1}{\rm Mpc}$, we do not find significant signals for galaxy-ellipticity, ellipticity-ellipticity, or spin correlations at both redshifts. Below $1 \; h^{-1}{\rm Mpc}$, a weak deviation from zero is seen at $z=1.47$, implying weak spin-filament correlations, but it is not verified by the direct comparison between angles of spins and filaments.
The linear alignment model fit yields the amplitude parameter $A_{\rm NLA}=1.38\pm2.32$ at $z=1.19$ and $0.45\pm2.09$ at $z=1.47$ ($95\%$ confidence levels).
We discuss various observational and physical origins that affect the search for alignments of disk galaxies at high redshifts.
\end{abstract}

\section{Introduction}
 \label{sec:intro}
The orientation of galaxies, derived from astronomical imaging data, is a valuable property as it delivers cosmological information through strong/weak gravitational lensing and provides insights into morphological evolution of galaxies \citep{Hoekstra:2008,Benson:2010,Treu:2010,Conselice:2014}. It has been established by numerous studies that the shape of galaxies, especially that for red populations, is not randomly oriented but tends to be coherent as a result of tidal interactions with the density field that they are embedded in \citep{Hirata:2004,Joachimi:2011,Okumura:2009,Singh:2015,Tonegawa:2022}.
Studies of such galaxy intrinsic alignments (IAs) have become important in multiple aspects, i.e., for (1) weak-lensing contamination, (2) galaxy formation and evolution, and (3) cosmological applications. First, the coherence of the galaxy orientation causes additional contribution to the signal of weak-lensing statistics by $\sim 10\%$ \citep{Troxel:2015}, leading to a systematic error on the cosmological inference if not properly understood and accounted for. Second, IA provides useful insights into the processes of galaxy formation and evolution that change galaxy shapes. Such processes include not only the interplay with the large-scale structures, but also mergers, stellar/AGN feedbacks, and accretions from the surrounding environments \citep{Velliscig:2015,Tenneti:2017,Bate:2020,Soussana:2020}. Cosmological-scale hydrodynamical simulations in $\gtrsim$ a few hundreds $h^{-1}{\rm Mpc}$, which includes baryonic physics on top of the gravitational evolution, are being actively used to study how IA depends on different galaxy properties and surrounding environments \citep{Dubois:2014,Welker:2014,Hilbert:2017,Bhowmick:2020,Samuroff:2021,Zjupa:2022,Xu:202311}. Third, due to the linear proportionality of IA to the density fluctuations \citep{Catelan:2001,Hirata:2004}, IA itself can be used as a probe for the underlying density fields. Along with the traditional probes using density fields \citep{Eisenstein:2005,Guzzo:2008}, baryon acoustic oscillations and redshift space distortions of IA in two-point statistics can be detected and utilized to constrain cosmology \citep{Chisari:2013,Okumura:2019,Okumura:2023,Xu:2023N,Dompseler:2023}. Primordial universe could even be studied through IA \citep{Schmidt:2014,Schmidt:2015,Matteo:2020,Shiraishi:2023,Okumura:2024,Saga:2024}. 
Understanding the degree of the strength of IA across various galaxy types and redshifts is therefore important.

The ellipticity correlation function is the primary quantity to study IAs. 
Its amplitude is known to significantly depend on galaxy types. Early-type galaxies have consistently exhibited positive signals towards overdensities, meaning that the major axis is preferentially oriented along the density fluctuation $\delta(\boldsymbol{x})$ \citep{Mandelbaum:2006,Hirata:2007,Okumura:2009b,Singh:2015,Johnston:2021,Tonegawa:2022}, and the tidal stretching model \citep{Catelan:2001, Hirata:2004} successfully explained the signals as a function of pair separation with a single amplitude parameter. On the other hand, no significant signal has been detected for late-type galaxies \citep{Hirata:2007,Mandelbaum:2011,Tonegawa:2018,Johnston:2021,Samuroff:2023,Xu:202309,Peters}, likely because late-type galaxies are more subject to the tidal torquing mechanism than to the tidal stretching \citep{Catelan:2001,Hirata:2004}. If this is true, the signal will be much more difficult to detect because the tidal torquing is a higher-order effect of the density fluctuation \citep{Schafer:2012,Blazek:2019}. A notable feature of the tidal torquing model is that it predicts null density-ellipticity correlations but positive ellipticity-ellipticity correlations for the Gaussian density field. Further observational data will be needed to verify whether the tidal torquing model is appropriate to explain the IA of late-type galaxies.

Using hydrodynamical simulations, several studies showed non-zero IA signals for late-type galaxies with opposite signs (i.e., perpendicular and parallel alignments) while others did not \citep{Chisari:2015,Tenneti:2016,Codis:2018,Veena:2019,Kraljic:2020,Shi:2021,Delgado:2023}, probably due to different implementations of baryonic physics. New theoretical approaches to IA are being developed (TATT: \cite{Blazek:2019} , halo model: \cite{Fortuna:2021}, EFT: \cite{Vlah:2020,Bakx:2023}, LPT:
\cite{Maion:2024,Chen:2024}) and tested against observation and simulation data \citep{Troxel:2018,Samuroff:2021}. More observational studies will help train both IA theories and the recipes of the hydrodynamical simulations. Compared to red galaxies, there are a handful of observational studies for blue galaxies, especially at high redshifts. More sample variations on galaxy properties and redshift would be desirable.

Another quantity that is relevant to late-type galaxies is the spin because the disks are rotation-supported. The spin can be estimated by the galaxy's position angle (PA) and axis ratio. With the thin-disk approximation, the axis ratio represents the radial component of the spin up to the sign degeneracy. The PA then expresses the rotation on the celestial sphere, responding to the angular component of the spin.
In the linear regime, the spin correlation function is proportional to the square of the matter correlation function, $\delta_L^4$ \citep{Lee:2011}.
The spin correlation function has been measured in the local universe \citep{Pen:2000,Lee:2011,Koo:2018}. Also, \citet{Motloch:2021} found a tentative correlation between the galaxy spin of the Galaxy Zoo sample and the initial condition of the Universe. Using the Horizon Run 5 hydrodynamic simulation \citep{Lee:2021}, \citet{Park:2022} found that the galaxy spin at $z=6$ is also correlated to the initial condition. 

If the tidal torquing is the primary mechanism for the alignments of late-type galaxies, the II correlation function is expected to be non-zero, as well as the spin correlation. However, observations so far \citep{Mandelbaum:2011,Tonegawa:2018,Johnston:2021} did not find any II signal as opposed to the detections of spin correlations above. The difference may be attributed to different properties of the sample \citep{Joachimi:2015}, different scales probed, and/or the use of different statistical quantities. Otherwise, the size and quality of observed shape samples currently may not be sufficient to reach the detectability for the weak signals that are seen in simulations (e.g., \cite{Dubois:2014}).

In this study, we use the up-to-date imaging data sets taken by the Hyper Suprime-Cam (HSC; \cite{Aihara:2019,Aihara:2022}) mounted on the Subaru Telescope, to measure the ellipticity two-point correlation functions for late-type galaxies at $z=1.19$ and $z=1.47$. 
We will also measure spin correlation function from the identical sample, in an attempt to disentangle the origins of the different indications between the II and spin correlations above.
The use of narrow-band observations \citep{Hayashi:2020} enabled us to identify [OII]-emitting galaxies at these redshifts effectively. 
The clustering properties of the [OII] sample have been extensively studied by \citet{Okumura:2021}.

This paper is organized as follows. In section \ref{section:theory}, we review the theory of the tidal torquing on the galaxy shape and spin alignments.
In section \ref{section:data}, we describe the sample used and in section \ref{section:measurement}, we explain the measurements of the observables. These are followed by the results in section \ref{section:results}. We conclude our work in section \ref{section:discussion}. The robustness of the measurements and model fitting is discussed in Appendix \ref{section:systematics}.

\section{Theory of Spin Alignments}\label{section:theory}
\subsection{Alignments from Tidal Torquing}
The internal structure of a galaxy is mainly determined by the surrounding tidal fields, which are created by the density contrast of the large-scale structures.
The tidal force is the difference of gravitational force across an object and has an impact on the internal structure of an object.
There are two types of effects from the tidal field: tidal stretching and tidal torquing.

The tidal stretching mechanism is considered to explain the ellipticity of early-type galaxies \citep{Catelan:2001,Hirata:2004}. 
Tidal torquing is another form of the deformation process of galaxies. For disky and rotation-supported galaxies, it is more natural to consider the angular momentum.  When the eigenvectors of the tidal field (denoted by a tensor ${\bf T}=\partial_i \partial_j \Phi$ with the gravitational potential $\Phi$) and the inertia tensor of galaxies (denoted by a tensor ${\bf I}$) are different, it exerts a torque to galaxies, leading to an angular momentum $\boldsymbol{L}$ \citep{White:1984},
\begin{eqnarray}\label{equation:J}
L_i \propto \epsilon_{ijk}I_{jl}T_{lk},
\end{eqnarray}
where $\epsilon_{ijk}$ is the Levi-Civita symbol.
The direction of the resulting spin will govern the observed ellipticity $\gamma^{\rm I}$ through the inclination angle as,
\begin{eqnarray}
\gamma^{\rm I} \propto \frac{1-\hat{L}^2_z}{1+\hat{L}^2_z},
\end{eqnarray}
where the $z$-axis is taken as the line of sight and the hat denotes unit vectors. 
Assuming 
that the inertia tensor is random and isotropic, the two-component ellipticity is given by \citep{Mackey:2002,Hirata:2004}
\begin{eqnarray}
(\gamma_1^{\rm I},\ \gamma_2^{\rm I})=C_2 \left(\tilde{T}
_{x\mu}^2-\tilde{T}_{y\mu}^2,\ 2\tilde{T}_{x\mu}\tilde{T}_{y\mu}\right),
\end{eqnarray}
where $\tilde{T}_{ij}=T_{ij}-\delta_{ij}{\rm Tr}({\bf T})/3$.
The auto power spectrum of $\gamma^{\rm I}$ is \citep{Hirata:2004}
\begin{eqnarray}
P_{\rm II}(k)=2\frac{C_2^2 \bar{\rho}^4 a^8}{\bar{D}^4} \int[h_E(\hat{\bf k}_1,\hat{\bf k}_2)]^2 \frac{P_{\delta\delta}^{\rm lin}(k_1)P_{\delta\delta}^{\rm lin}(k_2)}{(2\pi)^3} d^3k_1,
\end{eqnarray}
where ${\bf k}_2={\bf k}-{\bf k}_1$, $P_{\delta\delta}^{\rm lin}$ is the linear matter power spectrum, $a$ is the scale factor, $\bar{\rho}$ is the mean matter density, $\bar{D}(z)=(1+z)D(z)$ and $D$ is the growth factor normalized as $D(0)=1$, and $C_2$ is a free parameter specifying the overall amplitude of the shape correlation. This expression includes a kernel $h_E({\bf k},{\bf k'})$, which can also be found in calculations of \citet{Mackey:2002} in a similar manner. The II power spectrum is quadratic about the density two-point function $P(k)$, implying weak signals as we go to large scales.

To measure IAs, the ellipticity correlation function is often used. It is expressed as \citep{Crittenden:2001}
\begin{eqnarray}
\xi_{++}(r) \simeq a_T^2\alpha^2\frac{\xi_{\delta\delta}^2(r)}{\xi_{\delta\delta}^2(0)},
\end{eqnarray}
where $\alpha$ represents the thickness of galaxies, $a_T$ is a parameter related to the correlation between ${\bf T}$ and ${\bf I}$ (see the next subsection) and $\xi_{\delta\delta}$ is the matter two-point overdensity correlation function. Here, the subscript $+$ denotes the ellipticity component radial to overdensities.
The equation shows the proportionality between angular momentum-induced shape correlation and the underlying matter correlation function. The misalignments between galaxies and dark matter components will also affect the amplitude parameter $a_T$. Unlike the tidal stretching mechanism, tidal torquing predicts zero correlation between density fields (sampled by galaxies) and ellipticity, i.e. $\xi_{g+}(r)=0$ \citep{Mandelbaum:2011}. This difference is useful for constraining the alignment mechanism of galaxies from observations.

\subsection{Galaxy Spins}
The galaxy spin is generated by misalignments between the inertia of protogalaxies and the tidal field (equation (\ref{equation:J})).
If the eigenvectors of ${\bf T}$ are perfectly aligned with ${\bf I}$, the spin is not generated, but we generally cannot expect this because the formation of the protogalaxy involves baryon physics and nonlinear interactions such as mergers.
Still, as the protogalaxies are formed within the large-scale structure, ${\bf I}$ and ${\bf T}$ should be correlated to some degree.
\citet{Lee:2000} proposed a parametric model of the correlation of the unit spin vector, 
\begin{eqnarray}
\langle \hat{L}_i \hat{L}_j \rangle = \frac{1+a_T}{3}\delta_{ij} -a_T \hat{T}_{ik} \hat{T}_{kj},
\end{eqnarray}\label{equation:spin}
where $\hat{\bf T}$ is the unit tidal shear tensor ($\hat{\bf T} = \tilde{\bf T}/|{\bf T}|$). The parameter $a_T$ takes a value $0<a_T<3/5$ and quantifies the strength of the correlation. This leads to the expression of the spin correlation function
\begin{eqnarray}\label{equation:eta}
\eta(r) \simeq \frac{a_T^2}{6} \frac{\xi^2_{\delta\delta}(r;R)}{\xi_{\delta\delta}^2(0;R)},
\end{eqnarray}
where
\begin{eqnarray}
\xi_{\delta\delta}(r;R) = \frac{1}{2\pi^2}\int k^2W^2(kR)P_{\delta\delta}(k)j_0(kr) dk ,
\end{eqnarray}
with $W(kR)=\exp(-k^2R^2/2)$ and $j_0(kr)$ being the spherical Bessel function of $0$-th order. According to \citet{Lee:2011}, we take the smoothing scale $R= 1\; {\rm Mpc}/h$ when we fit the measurements.

\section{Data}\label{section:data}
The parent data for the [OII] sample is the second public data release (PDR2) of the HSC Subaru Strategic Program (SSP; \cite{Aihara:2019}). HSC has a wide field of view ($1.5\ {\rm deg}$) owing to the use of the primary focus of the Subaru Telescope. The survey obtained $grizy$ broad-band photometric data for three layers (Wide, Deep, and Ultra-Deep) with different survey depths and areas. For Deep and Ultra-deep layers, various narrow-band data is also available. 
\citet{Hayashi:2020} constructed [OII]-emitting galaxy samples over $16\ {\rm deg}^2$ on the sky, located at COSMOS, DEEP2, ELAIS-N, and SXDS regions, by utilizing the narrow-band information from NB816 (corresponding to $z=1.19$) and NB921 ($z=1.47$).
The detailed selection procedure for the [OII] sample can be found in \citet{Hayashi:2020} and \citet{Okumura:2021}. The clustering analysis of the sample has been performed by \citet{Okumura:2021}.
To ensure high completeness, we additionally impose conditions on the [OII] flux $\ge 3\times 10^{-17} {\rm erg s}^{-1} {\rm cm}^{-2}$ and the cmodel magnitude of each narrow-band $\le 23.5$. The resulting galaxy numbers in the four sky regions are $(544,\;2339,\;2722,\;2186)$ for NB816 and $(2987,\;1976,\;2903,\;1174)$ for NB921.

It is known that the background subtraction is slightly incomplete in the HSC PDR2 \citep{Aihara:2019}. Because such residual fluxes can cause systematics of shape measurements, we instead use the PDR3 data for shape measurements. The details of improvements on sky subtraction can be found in Sections 3.5 and 3.6 of \citet{Aihara:2022}. The two catalogs, PDR2 and PDR3, are cross-matched, and $>99\%$ of the [OII] galaxies had a counterpart within $0.2$ arcsec in PDR3.
The $5\sigma$ limiting magnitude for PDR3 is $(g,r,i,z,y)=(27.3,26.9,26.7,26.3,25.3)$ \citep{Aihara:2022}. For our sample, the median seeing is $(g,r,i,z,y)=(0.82,0.77,0.65,0.66,0.70)\ {\rm arcsec}$; $izy$ bands have better observing conditions than $gr$.
The median of stellar masses, which are estimated with the broad-band magnitudes and available from the data release website\footnote{https://hsc-release.mtk.nao.ac.jp/datasearch/}, is $4.3\times10^9 M_{\odot}$ ($z=1.19$) and $5.2\times 10^9 M_{\odot}$ ($z=1.47$). The majority of the objects are less than the transition mass $3\times10^{10} M_{\odot}$  \citep{Dubois:2014}, where the spin direction of galaxies relative to the nearest filament flips. 
We use the $z$-band information for shape measurements because we expect that the stellar component is best represented in the rest-frame $r$-band, which is redshifted to $z$-band at $z\sim1$. However, unlike early-type galaxies with $4000\; {\rm \AA}$ breaks, late-type galaxies are bright at shorter wavelengths as well. Therefore, we also present measurements using $i$ and $y$ bands, which have comparable seeing conditions to $z$-band (Appendix \ref{section:systematics}). 

The adaptive moments \citep{Bernstein:2002} of the second order, $I_{11}$, $I_{12}$, and $I_{22}$ are provided and used to obtain the two-component ellipticity,
\begin{eqnarray}
 e_1 &=& \frac{I_{11}-I_{22}}{I_{11}+I_{22}},\\
 e_2 &=&  \frac{2I_{12}}{I_{11}+I_{22}}.
\end{eqnarray}
These quantities are also provided for the point spread function (PSF) at the positions of each galaxy. We also calculate the ellipticity $(e_{1,{\rm PSF}}, e_{2, {\rm PSF}})$ to correct the ellipticity as 
\begin{eqnarray}
 {e}_{i, {\rm cor}} = \frac{e_i - Re_{i,{\rm PSF}}}{1-R},
\end{eqnarray}
where $R=e_{\rm PSF}/e$ \citep{Kaiser:1995}.
We remove objects with $e>0.8$ because they are not reliable.
As presented in Appendix \ref{section:systematics}, our results do not change when we use a more sophisticated shape catalog \citep{Li:2022}.

\begin{figure}
\includegraphics[width=0.47\textwidth]{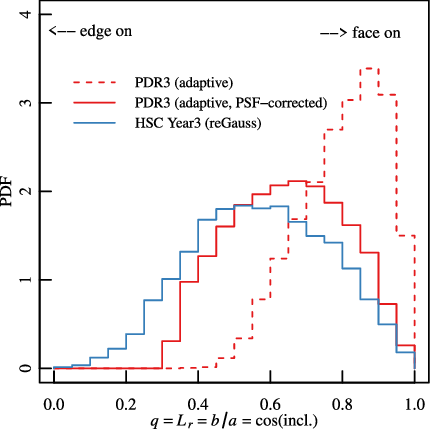}
\caption{Distribution of the axis ratio of the sample. Red lines are the shape measurements based on PDR3 data (dashed: raw, solid: PSF-corrected) . The blue line is drawn from the HSC Y3 shear catalog.}\label{figure:q}
\end{figure}

Figure \ref{figure:q} shows the distribution of the axis ratio $q$, which is related to $e$ through $e=\frac{1-q^2}{1+q^2}$. The correction for PSF reduces $q$ because PSF rendered objects rounder when objects are observed. Unfortunately, as the fourth-order moment is not provided in the PDR2 and PDR3, we only correct for PSF up to the second order. We also show the distribution of $q$ derived from the HSC Year 3 (Y3) shape data \citep{Li:2022}, in which the reGaussian method is used to estimate $e$ \citep{Hirata:2003,Rowe:2015}. The Y3 catalog exhibits a stronger correction of $q$. Yet, we find that the distributions deviate from the uniform distribution that is expected in the ideal thin-disk approximation.  This may be a common feature of observations; galaxies with high $e$ tend to be small projected on the image, making themselves harder to detect. Also, the existence of bulges can distort the distribution of projected axis ratio \citep{Zhang:2015} by increasing $q$. When we recover the radial component of the spin, lacking high $e$ (low $q$) preferentially selects galaxies with high $L_r$, which potentially causes a systematic offset of $\eta(r)$. For the spin correlation function, we correct for this effect and potentially incomplete removal of PSF by subtracting the ``shuffled'' measurements.

Random catalogs corresponding to the data sample are needed to measure the correlation functions (Section \ref{section:measurement}). We use a set of random points that were used in \citet{Okumura:2021}. It is based on the  value-added product of the HSC public data, in which $\sim 100\; {\rm arcmin}^2$ random points are distributed. The same bright star masks and flags as used in the sample selection for [OII] emitters \citep{Hayashi:2020} were applied to ensure that the random points traced the galaxy sample except for clustering.

\section{Measurements}\label{section:measurement}
\subsection{Alignments}
As our sample is limited to a relatively narrow range of distance due to the use of narrow bands, it is appropriate to use the angular correlation function $w_{AB}(\theta)$. It is related to the 3-D counterpart $\xi_{AB}(\boldsymbol{r})$, with the subscript $AB$ being $g+$, $g\times$, $++$, $\times\times$, or $+\times$,
\begin{eqnarray}
w_{AB}(\theta)= \int\int p(\chi_1)p(\chi_2)\xi_{AB}(r_p, r_\pi) d\chi_1 d\chi_2,
\end{eqnarray}
where $p(\chi)$ is the normalized distribution of comoving distances $\chi$ of the sample, $r_p=\frac{1}{2}(\chi_1+\chi_2) \theta$, $r_\pi=\chi_2 - \chi_1$, and $(r_p, r_\pi)$ are related to $(r, \mu)$ through $r^2 = r_p^2 + r_\pi^2$ and $\mu = r_\pi/r$. 
To obtain the redshift distribution of our sample, we use the filter transmission curves of NB816 and NB921\footnote{https://subarutelescope.org//Observing/Instruments/HSC/sensitivity.html}, by converting the wavelength to redshift for [OII]$
\lambda 3727$. According to \citet{Hayashi:2015}, the redshift distribution of the sample in the narrow-band survey is expected to trace the filter transmission curve. It is more realistic and accurate than adopting a simple top-hat radial function \citep{Okumura:2021}, since the edges of the filters do not capture emission lines fully, reducing the detection rate.

The galaxy-intrinsic alignment (gI) correlation function is measured by the modified Landy-Szalay estimator \citep{Mandelbaum:2006},
\begin{eqnarray}
 w_{g+}(\theta) &=& \frac{S^+D-S^+R}{RR}\\
 w_{g\times}(\theta) &=& \frac{S^\times D-S^\times R}{RR},
\end{eqnarray}
where '$S$', '$D$', and '$R$' mean the shape, density, and random samples. Each term is calculated on a pair-count basis,
\begin{eqnarray}
S^{+}D = \sum_{i\in S, j\in D} \frac{e_{+}(j|i)}{2\mathcal{R}},
\end{eqnarray}
where $e_{+}(j|i)$ and $e_{\times}(j|i)$ are the two-component ellipticity along the line joining point $i$ and $j$; $e_{+}(j|i)=e\cos{2\theta_{ji}}$ and $e_{\times}(j|i)=e\sin{2\theta_{ji}}$, with $\theta_{ji}$ being the angle between the joining line and the reference axis. $S^+R$ is obtained by replacing $D$ with $R$ and $RR$ is the pair-count of the random points. The factor $\mathcal{R}=1-\langle e_{\rm rms} \rangle^2$ is the shear responsivity \citep{Kaiser:1995} and is $\sim 0.8$ for our sample.

Similarly, the ellipticity-ellipticity (II) correlation functions are measured as 
\begin{eqnarray}
w_{++}(\theta) &=& \frac{S^+ S^+}{RR}\\
w_{\times \times}(\theta) &=& \frac{S^\times S^\times}{RR},\\
w_{+\times}(\theta) &=& \frac{S^+ S^\times}{RR},
\end{eqnarray}
where
\begin{eqnarray}\label{equation:S+S+}
S^{+}S^{+} = \sum_{i\in S, j\in S} \frac{e_{+}(j|i)e_{+}(i|j)}{4\mathcal{R}^2}
\end{eqnarray}
and $S^{\times}S^{\times}$ and $S^{+}S^{\times}$ are defined likewise. The signals of $w_{g\times}(\theta)$ and $w_{+\times}(\theta)$ are expected to be zero due to parity serving as diagnostics for systematics. The terms in the numerator, $SD$, $SR$, and $SS$ are rescaled to $RR$ to account for the difference of the mean density between galaxies and randoms. 

We take nine logarithmic bins in $0.004<\theta<1\; {\rm deg}$, which roughly corresponds to $0.2$ to $50\; h^{-1}{\rm cMpc}$ at the redshifts of our samples. We employ the jackknife resampling to estimate the error and covariance matrix of the measurements. Each survey field is divided into $N_{\rm sub}$ subregions. The division number of each field (COSMOS, DEEP2, ELAIS, SXDS) is the same as done in \citet{Okumura:2021}; $N_{\rm sub}=[9,\;35,\;34,\;37]$ at $z=1.19$ and $[41,\;35,\;34,\;9]$ at $z=1.47$ so that all subregions have similar areas. For each redshift (labelled by $l$) and field (labelled by $m$), $N_{\rm sub}^{l,m}$ jackknife realizations are performed excluding $k$-th subregion ($1\leq k \leq N_{\rm sub}^{l,m})$, giving a covariance matrix
\begin{eqnarray}
&&{\rm Cov}\left[X^{l,m}(\theta_i), X^{l,m}(\theta_j)\right] \nonumber \\
&&=\frac{N_{\rm sub}^{l,m}-1}{N_{\rm sub}^{l,m}}\sum_{k=1}^{N_{\rm sub}^{l,m}}\left[X_{(k)}^{l,m}(\theta_i)-\bar{X}^{l,m}(\theta_i)\right]\left[X_{(k)}^{l,m}(\theta_j)-\bar{X}^{l,m}(\theta_j)\right] \nonumber \\ 
\end{eqnarray}
where $X_{(k)}$ is the $k$-th realization for quantity $X$ and $\bar{X}$ is the mean of the $N_{\rm sub}^{l,m}$ realizations. Here, $X$ is $w_{g+}(\theta)$, $w_{g\times}(\theta)$, $w_{++}(\theta)$, $w_{\times\times}(\theta)$, $w_{+\times}(\theta)$, or $\eta(\theta)$ (next subsection).
At each redshift, measurements in the four fields are merged into the combined $X$ and associated covariance \citep{Okumura:2021},
\begin{eqnarray}
X^{l}(\theta) = \frac{\sum_{m=1}^{4} \left[W^{l,m}(\theta)\right]^2 \; X^{l,m}(\theta)}{\sum_{m=1}^{4} \left[W^{l,m}(\theta)\right]^2},
\end{eqnarray}
\begin{eqnarray}
&&{\rm Cov}\left[X^{l}(\theta_i), X^{l}(\theta_j)\right] \nonumber \\
&&= \frac{\sum_{m=1}^{4} \left[W^{l,m}(\theta_i)\right]^2\left[W^{l,m}(\theta_j)\right]^2 \;{\rm Cov}\left[X^{l,m}(\theta_i), X^{l,m}(\theta_j)\right]}{\sum_{m=1}^{4}\left[W^{l,m}(\theta_i)\right]^2 \; \sum_{m=1}^{4}\left[W^{l,m}(\theta_j)\right]^2}, \nonumber \\
\end{eqnarray}
where $[W^{l,m}(\theta_i)]^2={\rm Var}[X^{l,m}(\theta_i)$] is the diagonal element of ${\rm Cov}\left[X^{l,m}(\theta_i), X^{l,m}(\theta_j)\right]$.


\subsection{Spin}
By assuming that spiral galaxies are thin disks and that the spin direction is perpendicular to the disk, the galaxy spin $(\hat{L}_r,\hat{L}_\vartheta,\hat{L}_\phi)$ can be estimated using the axis ratio $q=[(1-e)/(1+e)]^{1/2}$ and PA,

\begin{eqnarray}
\hat{L}_r&=&\pm \cos{\zeta}\\
\hat{L}_\vartheta&=& (1-\cos^2{\zeta})^{1/2} \sin{\rm PA}\\
\hat{L}_\phi&=&(1-\cos^2{\zeta})^{1/2} \cos{\rm PA},
\end{eqnarray}
where, following \citet{Lee:2011}, we apply a small correction for the breakdown of the thin-disk approximation \citep{Haynes:1984}:
\begin{eqnarray}
    \cos^2{\zeta}=\frac{q^2-p^2}{1-p^2}.  
\end{eqnarray}
Here, the local coordinate $(r,\vartheta, \phi)$ is put at the position of each galaxy $({\rm RA}, {\rm Dec})$ with $\vartheta=\pi/2 - {\rm Dec}$ and $\phi={\rm RA}$. 

We apply $p=0.1$ but have confirmed that using $p=0.2$ does not affect the measurements.
The spin in the local coordinates is transformed into that in global Cartesian coordinates as
\begin{eqnarray}
\hat{L}_{x}^{+}&=& \hat{L}_r^{+} \sin{\vartheta}\cos{\phi} + \hat{L}_\vartheta\cos{\vartheta}\cos{\phi}-\hat{L}_{\phi}\sin{\phi}\\
\hat{L}_{y}^{+}&=& \hat{L}_r^{+} \sin{\vartheta}\sin{\phi} + \hat{L}_\vartheta\cos{\vartheta}\sin{\phi}+\hat{L}_{\phi}\cos{\phi}\\
\hat{L}_{z}^{+}&=&\hat{L}_r^{+} \cos{\vartheta}- \hat{L}_\vartheta \sin{\vartheta}.
\end{eqnarray}

Because we cannot determine the direction of spiral arms (i.e., clockwise or anticlockwise) without external information such as eye inspections and integral field spectroscopy, the radial component $\hat{L}_r$ is degenerate about its sign. Consequently, the spin in the Cartesian coordinate is either $\hat{\boldsymbol{L}}^{+}=(\hat{L}_{x}^{+},\hat{L}_{y}^{+},\hat{L}_{z}^{+})$ or $\hat{\boldsymbol{{L}}}^{-}=(\hat{L}_{x}^{-},\hat{L}_{y}^{-},\hat{L}_{z}^{-})$, where $\hat{\boldsymbol{{L}}}^{-}$ is obtained by replacing $\hat{L}_r^{+}$ with $\hat{L}_r^{-}$.
The spin correlation function is
\begin{eqnarray}
\eta(\theta)=\langle \hat{\boldsymbol{L}}({\bf \theta_0})\cdot\hat{\boldsymbol{L}}({\bf \theta_0}+{\bf \theta}) \rangle - \frac{1}{3},
\end{eqnarray}
where, considering the above ambiguity,

\begin{eqnarray}
&&\langle \hat{\boldsymbol{L}}({\bf \theta_0})\cdot\hat{\boldsymbol{L}}({\bf \theta_0}+{\bf \theta}) \rangle \nonumber \\
&&= \frac{1}{4}\left(\langle \hat{\boldsymbol{L}}^{+} \cdot \hat{\boldsymbol{L}}^{+}\rangle + \langle \hat{\boldsymbol{L}}^{+} \cdot \hat{\boldsymbol{L}}^{-}\rangle
+ \langle \hat{\boldsymbol{L}}^{-} \cdot \hat{\boldsymbol{L}}^{+}\rangle + \langle \hat{\boldsymbol{L}}^{-} \cdot \hat{\boldsymbol{L}}^{-}\rangle\right)
\end{eqnarray}

If $e$ has a systematic offset due to an incomplete correction for PSF, there will be an artificial correlation of the spin direction. To subtract this effect, we also calculate $\eta_{\rm shuf}(\theta)$ by randomly shuffling the spin direction within each FoV. The shuffling is performed $100$ times, each realization giving $\eta_{\rm shuf}^k(\theta)$ ($k=1,\;2,\;\dots,\;100)$, and then these are averaged to obtain $\eta_{\rm shuf}(\theta)$. Hereinafter, the presented $\eta(\theta)$ is the one after $\eta_{\rm shuf}(\theta)$ is subtracted.

\section{Results}\label{section:results}
\begin{figure*}
\includegraphics[width=0.97\textwidth]{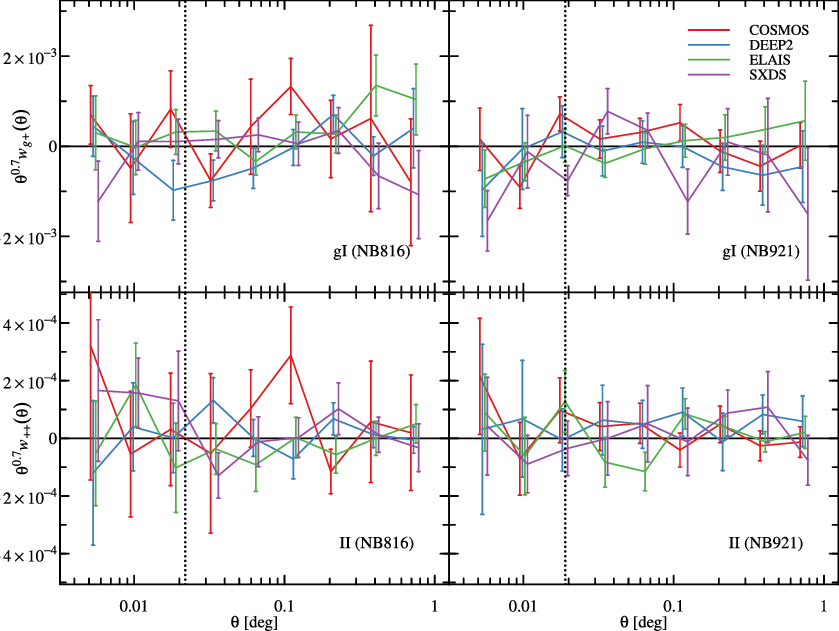}
\caption{The gI correlation function $w_{g+}(\theta)$ (top panels) and II correlation function $w_{++}(\theta)$ (bottom panels) as a function of separation angle $\theta$. The left panels are for $z=1.19$ and the right panels are for $z=1.47$. 
Black lines are the combined results and other colors are results for different field-of-views. The vertical dotted lines indicate the comoving separation of $1 \; {\rm Mpc}/h$.}\label{figure:IA_FoV}
\end{figure*}

\begin{figure*}
\includegraphics[width=0.97\textwidth]{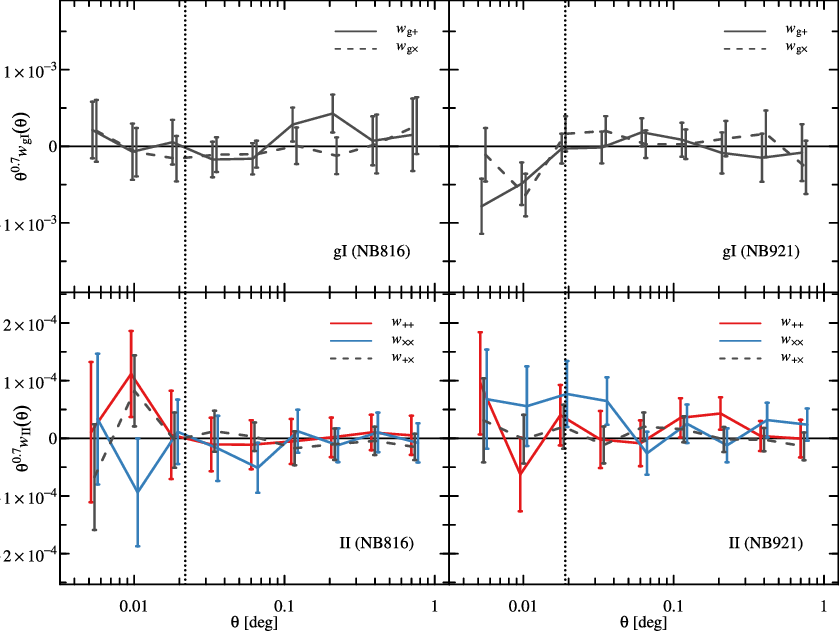}
\caption{The combined gI correlation functions $w_{g+}(\theta)$ and $w_{g\times}(\theta)$ (top panels) and the combined II correlation function $w_{++}(\theta)$, $w_{+\times}(\theta)$, and $w_{\times\times}(\theta)$ (bottom panels ) as a function of separation angle $\theta$. The left panels are for $z=1.19$ and the right panels are for $z=1.47$. $w_{g+}$ and $w_{+\times}$ can be used as diagnosis of systematics, as they are expected to be zero.}\label{figure:IA_all}
\end{figure*}

Figure \ref{figure:IA_FoV} shows the galaxy-ellipticity correlation function $w_{g+}(\theta)$ (top panels) and the ellipticity-ellipticity correlation function $w_{++}(\theta)$ (bottom panels). Measurements in each FoV are presented as colored lines. As expected, the COSMOS region of NB816 and the SXDS region of NB921 are noisier than other FoVs due to small number of galaxies. Otherwise, the measurements are consistent with one another within error bars. Therefore, we will only discuss the combined (i.e., FoV-averaged) results. The combined results of $w_{g+}(\theta)$, $w_{g\times}(\theta)$, $w_{++}(\theta)$, $w_{+\times}(\theta)$, and $w_{\times\times}(\theta)$ are shown in Figure \ref{figure:IA_all}. For $w_{g+}(\theta)$, we do not observe significant correlation signals at both $z=1.19$ and $z=1.47$ beyond $1\; h^{-1}{\rm Mpc}$. This is consistent with the prediction from the tidal torquing model; the gI correlation is the three-point function of the overdensity and its expectation value is zero on linear scales \citep{Catelan:2001}. In the non-linear regime at $z=1.47$, the combined result slightly falls below zero. With the smallest two bins, the non-linear alignment model fitting (see section \ref{section:discussion}) gives $\chi^2_{\rm min}=1.05$, while the null hypothesis ($A_{\rm NLA}=0$) yields $\chi^2=6.52$. The resulting $\Delta \chi^2 = 5.47$ indicates that the null hypothesis corresponds to a $\sim 2\sigma$ level deviation, and that the major axes tend to be perpendicular to the direction to the overdensity. As the spin and major axis of spiral galaxies tend to be perpendicular, negative $w_{g+}$ would imply that the spin direction is aligned with the line joining two galaxies, which is the direction of density fluctuations. Several studies report from hydrodynamical simulations \citep{Welker:2014,Dubois:2014,Shi:2021} that the spin of low-mass disk galaxies ($<10^{10} M_{\odot}$) tend to be parallel to the spine of the cosmic filaments. The phenomenon is caused by the angular momentum transfer from filaments during the smooth accretion, which is a non-linear effect.


For the II correlations, on large scales ($> 10\;h^{-1}{\rm Mpc}$) we do not observe signals, which is consistent with previous studies \citep{Mandelbaum:2011,Tonegawa:2018} finding no II signals for star-forming galaxies. While the tidal torque theory predicts non-zero II correlations \citep{Catelan:2001,Hirata:2004}, observational factors affect detectability, which will be discussed in the next section.
A recent simulation study reports negligible II correlations beyond $1 \;h^{-1}{\rm Mpc}$, even with $\sim 200,000$ spiral galaxies within a $500 \;h^{-1}{\rm Mpc}$-sized box \citep{Delgado:2023}. Considering that the tidal torquing is a weak effect compared to the tidal stretching for early-type galaxies and that the observational data suffer from additional noises such as PSF, it may be difficult to detect the imprints of the tidal torquing mechanism at large scales with the current data sets.
The tidal torquing itself is being challenged by several authors as the main driver of the alignments for spiral galaxies \citep{Samuroff:2021,Moon:2024}.

At $z=1.47$, a weak deviation from zero can be seen in $w_{\times\times}(\theta)$ below a few $h^{-1}{\rm Mpc}$. In fact, \citet{Delgado:2023} found marginal II signals in the one-halo regime, and this was not surprising as gI deviated from zero. At such scales, the main driver of the shape alignment is the non-linear gravitational interactions inside or in the vicinity of the cosmic web such as filaments \citep{Welker:2014,Dubois:2014} rather than the tidal torquing mechanism.
On the other hand, the signal is not seen at $z=1.2$ at all.
In general, as redshift decreases, the spin-filament alignment becomes weaker because of stochastic processes such as mergers \citep{Dubois:2014}.
However, the signal may be just affected by the statistical error, showing the apparent difference between the two redshifts. 

Figure \ref{figure:IA_all} also shows the cross component, $w_{g\times}(\theta)$ and $w_{+\times}(\theta)$.
This quantity can be used as a diagnostic of systematics, as it should be zero due to parity. Both quantities are consistently zero within error bars, indicating that our measurements do not suffer from systematics.

\begin{figure}
\includegraphics[width=0.47\textwidth]{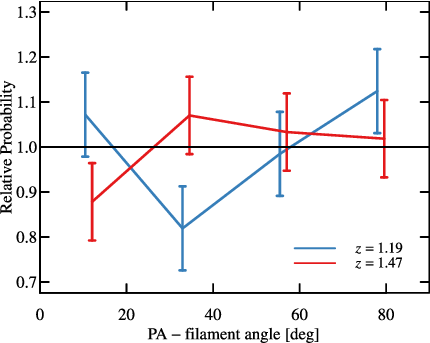}
\caption{The shape-filament alignment. For shape, those with $e>0.6$ and $0.005<D_{\rm nearest}<0.02 \;{\rm deg}$ are used. Error bars are from the Poisson statistics.}\label{figure:filaments}
\end{figure}

To further investigate whether the features observed in one-halo regime at $z=1.47$ are physical or not, we show in  Figure \ref{figure:filaments} the probability distribution of the angle between the major axis of the sample and the nearest filament segment on the 2-D (RA-Dec) plane. Here, the filaments are identified from the same [OII] sample using the DisPerSE code \citep{Sousbie:2011}, a structure identification algorithm based on the discrete Morse theory and persistent theory (an example of the detected filaments is given in the Appendix). Only edge-on galaxies with $e>0.6$  and $0.005<D_{\rm nearest}<0.02 \;{\rm deg}$ are used to produce Figure \ref{figure:filaments}. While there is a $\sim 2 \sigma$ signal in the top-right panel of  Figure \ref{figure:IA_all}, Figure \ref{figure:filaments} does not show excess at large angles between PAs and filaments, with $p$-value from the Kolmogorov-Smirnov test being $0.1365$ at $z=1.19$ and $0.4734$ at $z=1.47$ for the null hypothesis of the random orientation.
Considering that the negative $w_{g+}$ signals do not appear at $z=1.2$ in Figure \ref{figure:IA_all} and the low significance of shape-filament alignments at both redshifts in Figure \ref{figure:filaments}, we conclude that the signal in Figure \ref{figure:IA_all} is likely to be affected by statistical bias in our data set.

\citet{Dubois:2014} showed that the spin direction of spiral galaxies relative to the filaments became from perpendicular (high-mass end) to parallel (low-mass end) at a transition mass of $\sim 3\times 10^{10} \; M_{\odot}$.\footnote{According to \citet{Codis:2018}, the transition mass 
 for galaxies does not change significantly between $z=1.2$ and $z=1.5$. We assume that this holds true for our samples as well.}
To check for the possibility that our null signals are caused by the cancellation between the signals from the low-mass and high-mass subsamples, we also measured all of the above statistics, splitting the sample by the median stellar mass, finding null signals consistently (Appendix \ref{section:systematics}). This is likely because the majority of our galaxies have stellar masses below the transition mass of $10^{10.25-10.75} M_{\odot}$. Also, we do not find environment dependence by splitting the sample by the 2-D local density. 


\begin{figure*}
\includegraphics[width=0.97\textwidth]{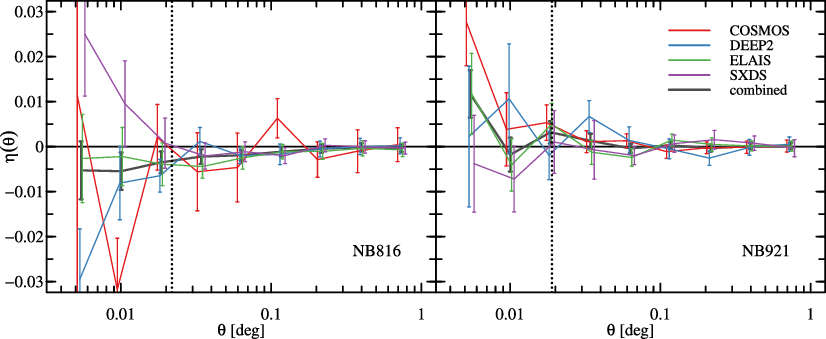}
\caption{The spin correlation function $\eta(\theta)$ as a function of separation angle $\theta$. The top panel is for $z=1.19$ and the bottom for $z=1.47$. 
Black lines are the combined results and other colors are for different FoVs.
}\label{figure:spin}
\end{figure*}

Figure \ref{figure:spin} shows the spin correlation function, $\eta(\theta)$.
There is a weak correlation for $z=1.47$ below $\theta=0.1 \; {\rm deg}$. 
The deviation from zero on scales below $1 \; h^{-1}{\rm Mpc}$ is in line with the gI and II correlation functions.
At larger scales, on the other hand, we do not see signals. As the signal predicted by the tidal-torque theory is proportional to $\delta^4$, the signal should be short-ranged if present \citep{Schafer:2012}, which explains the rapid decrease in signal as the separation increases. 
By fitting the theoretical prediction (equation (\ref{equation:eta})) beyond $1 \; h^{-1}{\rm Mpc}$, we obtain a constraint of $a_{T}=0.069^{+0.0548}_{-0.0456}$ ($z=1.47$). The theory of \citet{Lee:2000} prohibits negative $\eta(r)$, so the best-fit for $z=1.19$ is obviously $a_T=0$. The $\chi^2$ value for $a_T=0$ is $6.60/6$. 
We notice that the scale dependence of $\eta(\theta)$ is somewhat similar to $w_{++}(\theta)+w_{\times \times}(\theta)$. This implies that the spin and II correlation functions deliver similar information because the spin is reconstructed from the position angles and ellipticity too.


\section{Discussions}\label{section:discussion}
We have measured the ellipticity two-point correlation functions and the spin correlation function for [OII]-emitters at $z=1.19$ and $1.47$, selected by the narrow-band filters of the HSC imaging survey \citep{Hayashi:2020}.
Unlike early-type galaxies, the existence or absence of alignments of late-type galaxies is not conclusive either in observations \citep{Pen:2000,Lee:2011,Mandelbaum:2011,Tonegawa:2018,Koo:2018,Motloch:2021} and simulations \citep{Dubois:2014,Chisari:2015,Tenneti:2016,Veena:2019,Kraljic:2021,Zhang:2023}.
The reason for this discrepancy may be the difference of the sample used, and/or, the difference of the adopted statistical quantities, i.e., the shape correlation function and the spin correlation function.  We generally did not find significant signals either in the shape and spin correlation functions, except for the one-halo regime at $z=1.47$.

\begin{figure}
\includegraphics[width=0.47\textwidth]{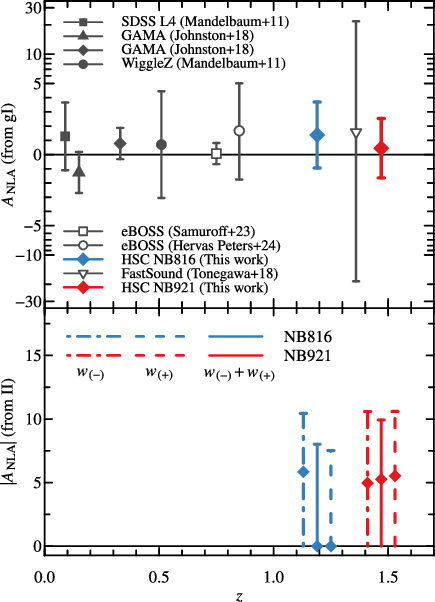}
\caption{Constraints on the amplitude of the NLA model fit, $A_{\rm NLA}$, as a function of redshift for late-type galaxies. (Top panel) Constraint from the galaxy-ellipticity correlation. The diamonds represent this study while other points are drawn from the literature. The error bars correspond to $2\sigma$. (Bottom Panel) Constraint from ellipticity-ellipticity correlation. 
Although different works use different conventions of the NLA model to obtain $A_{\rm NLA}$, we converted their results to match ours, specifically the one from \citet{Hirata:2010} (not \citet{Hirata:2004}) with $D(0)=1$. Note also that the $y$-axis of the top panel is logarithmic beyond $|A_{\rm NLA}|=5$.}\label{figure:A_NLA}
\end{figure}

To put new constraints on the tidal stretching mechanism on blue galaxies\footnote{While it is not clear to what extent the tidal stretching (not to be confused with torquing) work for blue galaxies, fitting with this stretching model has been done in the literature to put upper limits.}, we perform the linear alignment (LA) model fitting to $w_{g+}(\theta)$. The galaxy-intrinsic ellipticity power spectrum for the tidal stretching model is given by \citet{Hirata:2010,Okumura:2020}
\begin{eqnarray}\label{equation:NLA}
P_{g+}(\boldsymbol{k})=\frac{bC_1 \bar{\rho}}{\bar{D}(z)}a^2 
 (1-\mu_{\boldsymbol{k}}^2) P_{\delta \delta}(k),
\end{eqnarray}
where $b$ is the galaxy bias and $\mu_{\boldsymbol{k}}$ is the cosine between ${\boldsymbol{k}}$ and the line of sight. The amplitude parameter $C_1$ responds to all effects that change the alignment.
The use of nonlinear matter power spectrum $P_{\delta\delta}$ explains observations very well (NLA model; \cite{Bridle:2007}). According to \citet{Okumura:2021}, we fix $b=1.61$ ($z=1.19$) and $b=2.09$ ($z=1.47$) for our samples.
The $g+$ correlation function is obtained by Hankel transforming $P_{g+}$,
\begin{eqnarray}
    \xi_{g+}(r,\mu) = \frac{bC_1 \bar{\rho}}{\bar{D}(z)}a^2 (1-\mu^2)\int \frac{k^2dk}{2\pi^2}P_{\delta \delta}(k)j_2(kr),
\end{eqnarray}
where $\mu$ is the cosine between $\boldsymbol{r}$ and the line of sight.
We vary $A_{\rm NLA}=C_1 / [5\times10^{14} (h^2\;M_{\odot}{\rm Mpc}^{-3})^{-1}]$ to fit $w_{g+}(\theta)$ using the largest five $\theta$ bins (ensuring $> 3 h^{-1} {\rm Mpc}$). 

Figure \ref{figure:A_NLA} shows the $2\sigma$ constraint on $A_{\rm NLA}$, together with the results from previous works. We obtain $A_{\rm NLA} = 1.38\pm2.32$ ($z=1.19$) and $0.45\pm2.09$ ($z=1.47$); the positiveness at $z=1.19$ can be seen in Figure \ref{figure:IA_all}.
While all data points are consistent with zero within $2\sigma$, positive values are preferred in Figure \ref{figure:A_NLA}. This may reflect the tidal stretching mechanism on the bulge component \citep{Jagvaral:2022}, as interpreted as the tidal stretching + tidal torquing model \citep{Troxel:2018,Samuroff:2021}. In Appendix \ref{section:systematics}, we show the dependence of $A_{\rm NLA}$ for various conditions.
We also attempt to constrain $A_{\rm NLA}$ from the II correlations, by fitting $w_{(\pm)}(\theta) \equiv w_{++}(\theta) \pm w_{\times\times}(\theta)$ with the NLA model.
Expressions for the II power spectra can be found in \citet{Okumura:2020}. 
Due to the quadratic dependence, we can only constrain $A_{\rm NLA}^2$ from the II correlation. Without prior on the sign of $A_{\rm NLA}$\footnote{Early-type galaxies are well known to have positive $A_{\rm NLA}>0$, and it would be reasonable to set a prior to prohibit negative values: $p(A_{\rm NLA}<0)=0$.}, the PDF of $A_{\rm NLA}$ is symmetric (and generally bimodal) around $A_{\rm NLA}=0$ \citep{Bakx:2023}, giving the same absolute values for the lower and upper limits. Therefore, we only show $|A|$ in the bottom panel of Figure \ref{figure:A_NLA}. The constraints are $|A_{\rm NLA}(z=1.19)|<8.02$ and $|A_{\rm NLA}(z=1.47)|<9.93$.

While the signals obtained in this work were limited, we notice that the scale dependence of the II correlation function $w_{++}(\theta)+w_{\times\times}(\theta)$ and spin correlations $\eta(\theta)$ appeare to be similar (Figure \ref{figure:IA_all} and \ref{figure:spin}). 
By the fitting with power-law scaling of $\propto \alpha \theta^{-\beta}$ and marginalizing over $\alpha$, the slope is found to be consistent: $\beta=0.74\pm0.28$ (from II) and $0.60\pm0.35$ (from spin) at $z=1.47$. At $z=1.19$, $\beta=0.79\pm0.36$ (from II) and $0.40\pm0.23$ (from spin), which are still consistent, though less so, likely due to statistical reasons.
This consistency between the slopes from different quantities seems to be caused by the fact that both quantities are derived from the same information, i.e., the ellipticity and position angle. Also, the theory predicts the same scale dependence as $\xi_{\delta \delta}^2$ (Section \ref{section:theory}). Therefore, signals would be detected if the $S/N$ ratio is high, regardless of whether the II correlation or spin correlation is used.
For example, \citet{Pen:2000} found non-zero $\eta(\theta)$ but \citet{Mandelbaum:2011} and \citet{Tonegawa:2018} did not detect $w_{g+}(r_p)$.
Such discrepancy between II and spin correlations should be attributed to the difference in the sample properties such as galaxy type and redshift rather than statistical quantities.  

\citet{Dubois:2014} showed that the spin directions relative to filaments depend on stellar mass. This finding was observationally supported by \citet{Barsanti:2022}.
Within our sample, we did not find any dependence on stellar mass. We neither saw a dependence on the local density.

In this study, we used [OII]-emitters, which are expected to have strong star-forming activities. The morphology of such galaxies is likely to be late-type disk galaxies such as Sc and Sd.
However, as our sample selection was based solely on the spectral information, the morphological types of some galaxies can be irregular as well. Irregular galaxies often show a 
similar SED to spiral galaxies and can constitute a substantial fraction of galaxy demographics \citep{Gallagher:1984}. For irregular galaxies, we do not have enough knowledge about how the spin direction is determined by the large-scale structures because the turbulence is caused not only by gravitational effects but also by baryonic/stochastic processes such as interactions with other galaxies and unusual gas distributions. Naively, the shape turbulence would be random in direction, smearing the II and spin correlation functions. Also, the thin-disk approximation may not be valid for the irregulars. We have not put additional conditions to discriminate between these morphological types. If the fraction of irregulars is high, which likely happens at $z>1$, the true signal of spiral galaxies can be hidden by this contamination.
\citet{Lee:2011} and \citet{Motloch:2021} used the morphology information to build galaxy samples and detected the alignment signals, while \citet{Mandelbaum:2011} and \citet{Tonegawa:2018} did not. 
Several simulation projects detected spin alignments \citep{Shi:2021,Zjupa:2022}, and these works used the ratio of the rotational kinetic energy to the total kinetic energy, $\kappa_{\rm rot}$, to select rotation-dominated galaxies as disk galaxies. This criterion could have effectively removed irregular galaxies, leading to a marginal detection of the signals. Note that, the simulation is free from noises other than cosmic variance, which is responsible for the detectable $S/N$. Further investigation on whether irregular galaxies have strong alignments or not will be an interesting path to pursue. 

There are observational factors that can affect the measurements, especially at high redshifts.
First, the distribution of ellipticity (related to $q$ as $e=(1-q^2)/(1+q^2)$) deviates from the uniform distribution as shown in Figure \ref{figure:q}.  This is likely because galaxies with high $e$ tend to be small projected on the image, making themselves harder to detect, and because the existence of bulge components distorts the ellipticity. When we recover the radial component of the spin, lacking high $e$ (low $q$) systematically selects galaxies with high $L_r$, which potentially causes a systematic offset of $\eta(\theta)$. This is usually not a serious problem for IA of red galaxies, as $e$ only acts as weights.
Second, incomplete corrections for PSF potentially have a similar effect because PSF makes objects rounder (toward lower $e$). At $z>1$, the object size becomes comparable to the PSF size, and in such a case, precise PSF correction is more difficult. Third, \citet{Andrae:2011} noted that the kernel to estimate the ellipticity is important for disk galaxies. The use of second moment-based ellipticity may suffer from underestimation of $e$ compared to using the isophotes because it captures not only the disk features but also the galactic bulges.
These factors will become important at high redshift observations and take careful interpretations of the measurements.


\section*{Acknowledgements}
MT is supported by an appointment to the JRG Program at the APCTP through the Science and Technology Promotion Fund and Lottery Fund of the Korean Government, and was also supported by the Korean Local Governments in Gyeongsangbuk-do Province and Pohang City. MT was supported by the National Research Foundation of Korea (NRF) grant funded by the Korea government (MSIT) (2022R1F1A1064313).
TO acknowledges support from the National Science and Technology council under Grants No. NSTC 112-2112-M-001-034- and No. NSTC 113-2112-M-001-011-. 

The Hyper Suprime-Cam (HSC) collaboration includes the astronomical communities of Japan and Taiwan, and Princeton University.  The HSC instrumentation and software were developed by the National Astronomical Observatory of Japan (NAOJ), the Kavli Institute for the Physics and Mathematics of the Universe (Kavli IPMU), the University of Tokyo, the High Energy Accelerator Research Organization (KEK), the Academia Sinica Institute for Astronomy and Astrophysics in Taiwan (ASIAA), and Princeton University.  Funding was contributed by the FIRST program from the Japanese Cabinet Office, the Ministry of Education, Culture, Sports, Science and Technology (MEXT), the Japan Society for the Promotion of Science (JSPS), Japan Science and Technology Agency  (JST), the Toray Science  Foundation, NAOJ, Kavli IPMU, KEK, ASIAA, and Princeton University.

This paper is based on data collected at the Subaru Telescope and retrieved from the HSC data archive system, which is operated by Subaru Telescope and Astronomy Data Center (ADC) at NAOJ. Data analysis was in part carried out with the cooperation of Center for Computational Astrophysics (CfCA) at NAOJ.  We are honored and grateful for the opportunity of observing the Universe from Maunakea, which has the cultural, historical and natural significance in Hawaii.

This paper makes use of software developed for Vera C. Rubin Observatory. We thank the Rubin Observatory for making their code available as free software at http://pipelines.lsst.io/.


\appendix
\section{Measurements and constraints on $A_{\rm NLA}$ for various configurations}\label{section:systematics}

\begin{figure*}
\begin{center}
\includegraphics[width=0.8\textwidth]{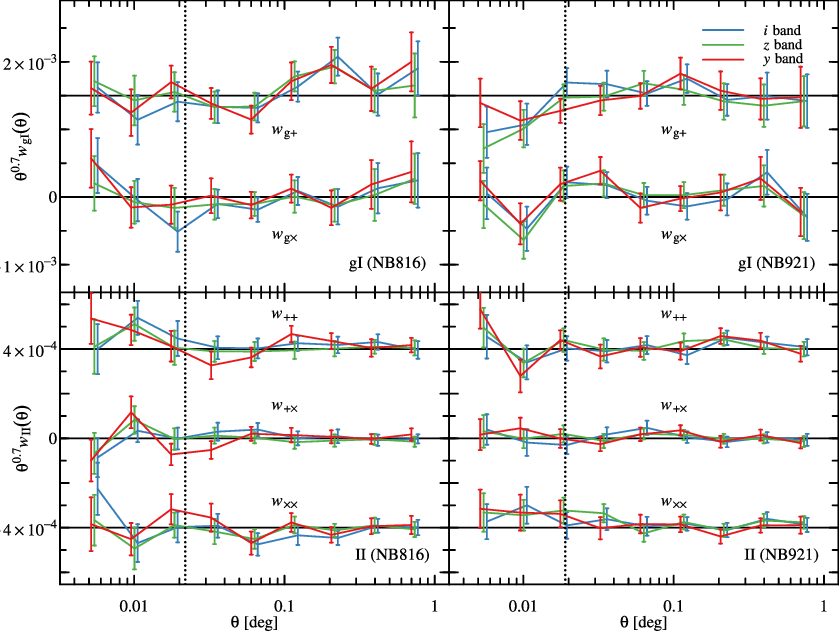}
\end{center}
\caption{Comparison of $w_{\rm gI}(\theta)$ (top panels) and $w_{\rm II}(\theta)$ (bottom panels) from different imaging filters. The left panels are for $z=1.19$ and the right panels are for $z=1.47$. All are derived from PDR3. For a viewing purpose, $w_{g+}$, $w_{++}$, and $w_{\times\times}$ are intentionally offset as indicated by solid horizontal lines. 
}\label{figure:filters}
\end{figure*}

\begin{figure*}
\begin{center}
\includegraphics[width=0.8\textwidth]{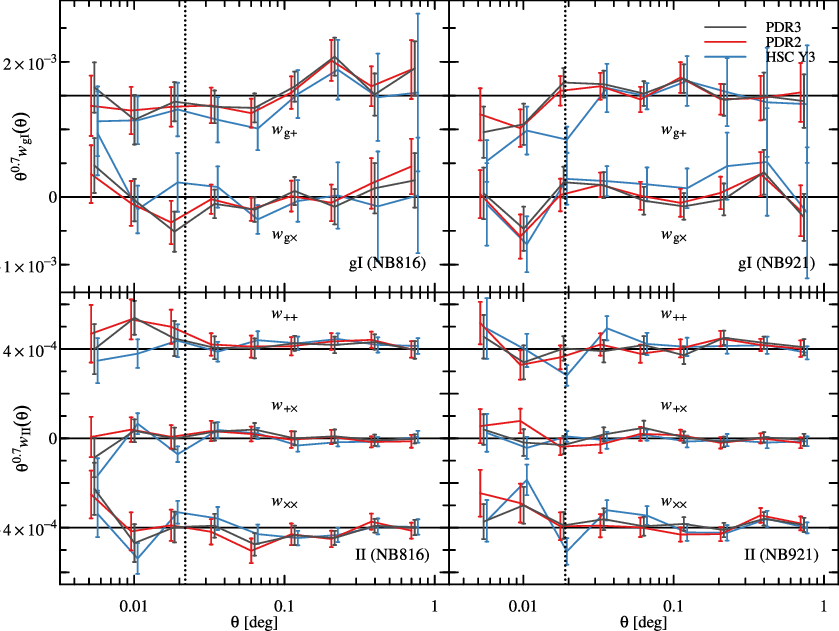}
\end{center}
\caption{Similar to figure \ref{figure:filters}, but the comparison between of different shape data (PDR2, PDR3, and Y3). The $i$-band information is used for all data sets due to the availability of Y3.}\label{figure:data}
\end{figure*}

We used the $z$-band imaging of the PDR3 data in the main text.
Here, we examine whether our results are affected by the choices of data and by physical properties of the sample.
Figure \ref{figure:filters} presents how the IA correlation functions depend on the filter used. The $i$ and $y$-band results are displayed on top of the $z$-band. These bands also correspond to the stellar emission in the rest frame. All measurements are consistent with each other.
Figure \ref{figure:data} show the different data set, PDR2 \citep{Aihara:2019} and the HSC Year 3 shape catalog \citep{Li:2022}. The residual sky subtraction is reported in PDR2 but the effect on the correlation functions is negligible. The Y3 catalog uses the reGaussianization method \citep{Hirata:2003} for shape estimation, and has less number of galaxies due to the quality cut. By cross matching with PDR3, the number is $[404, 1329, 0, 1585]$ and $[1874, 907, 0, 755]$ (the ELAIS region is not available). The general trend is similar to the PDR3 case, except for the small scale of $w_{++}$ at $z=1.19$. Overall, our reuslts are not affected by the filters or data sets.

\begin{figure*}
\begin{center}
\includegraphics[width=0.8\textwidth]{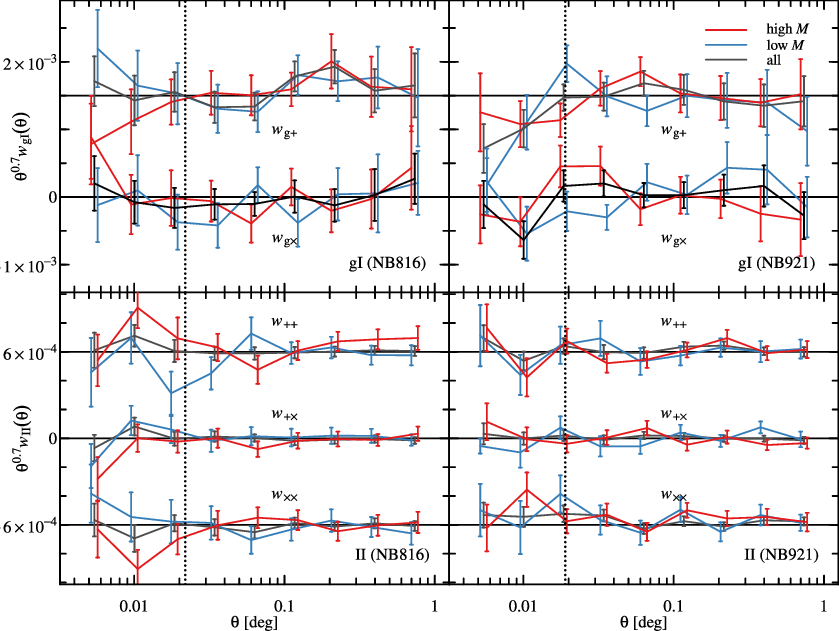}
\end{center}
\caption{Similar to figure \ref{figure:filters}, but the comparison between different subsamples divided by the stellar mass. All are obtained using $z$-band imaging data of PDR3 to be consistent with the main text.}\label{figure:mass}
\end{figure*}

\begin{figure*}
\begin{center}
\includegraphics[width=0.8\textwidth]{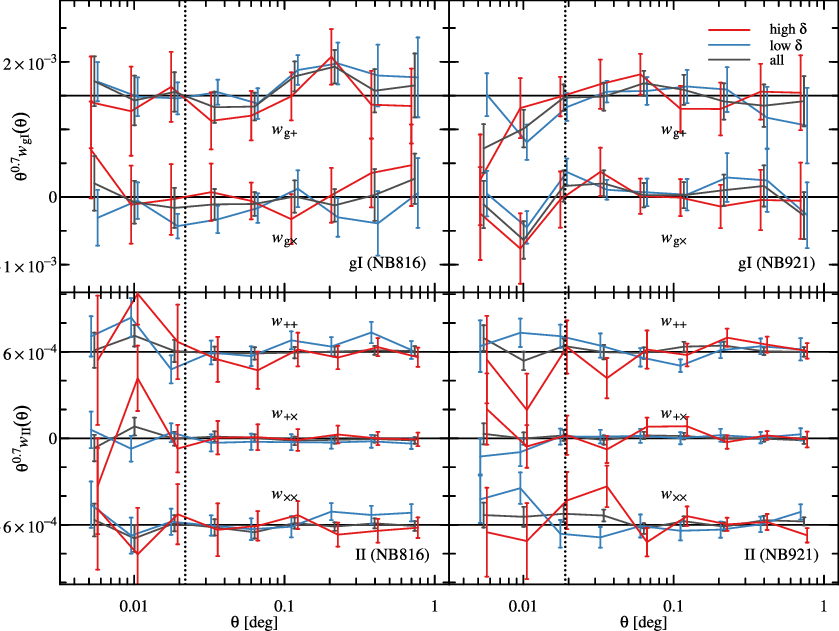}
\end{center}
\caption{Similar to figure \ref{figure:filters}, but the comparison between different subsamples divided by the 2-D local density, estimated by the five nearest neighboring galaxies, $\delta=5/(d_{\rm 5th})^2$. All are obtained using $z$-band imaging data of PDR3 to be consistent with the main text.}\label{figure:localdensity}
\end{figure*}

\begin{figure}
\includegraphics[width=0.47\textwidth]{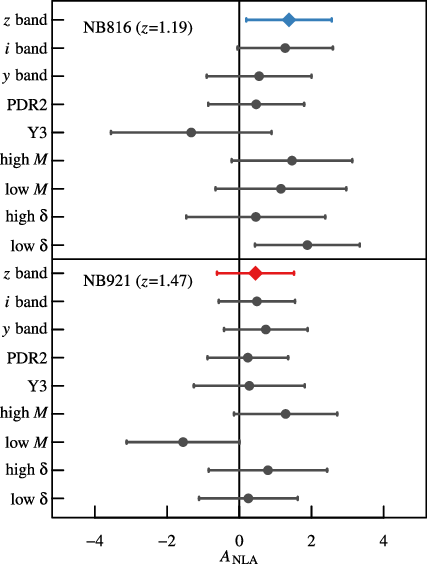}
\caption{The amplitude parameter $A_{\rm NLA}$ estimated from $w_{g+}(\theta)$ measurements with various conditions as presented in Figure \ref{figure:filters} to Figure \ref{figure:localdensity}. The fiducial results, shown as colored diamonds, are obtained using the $z$-band of the PDR3 data (see main text).}\label{figure:A_NLA_sys}
\end{figure}

\begin{figure}
\includegraphics[width=0.47\textwidth]{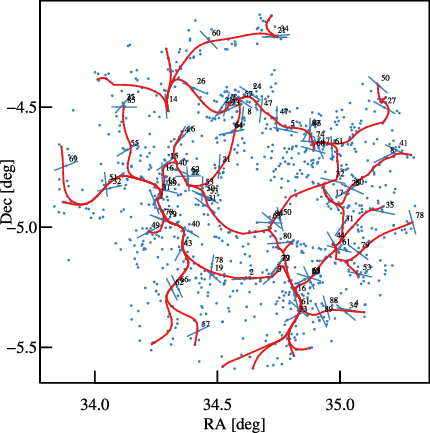}
\caption{An example of the 2D filamentary structure detected by DisPerSE ($2\sigma$) on the DEEP2 field at $z=1.47$. Blue points are the whole [OII] sample. Those with arrows (the direction of major axis) satisfy $0.005 < D_{\rm nearest} < 0.02 \;{\rm deg}$ and $0.6<e<0.8$, from which Figure \ref{figure:filaments} is produced. The numbers near arrows indicate the angle between the major axis and the nearest filament spline.}\label{figure:disperse}
\end{figure}

Figure \ref{figure:mass} and \ref{figure:localdensity} show the dependence on the sample properties. We split the PDR3 sample into two using the median of stellar masses (Figure \ref{figure:mass}), which are estimated with the broad-band magnitudes and available from the HSC data release website. This split is motivated by findings of the flipping of spin-filament alignments \citep{Dubois:2014}. If the mass range of our sample trespasses the transition mass, the cancellation between galaxies perpendicular and parallel to the filaments occurs and the alignment signal would be weaken. In Figure \ref{figure:mass}, we do not find a dependence on stellar mass, indicating that the null detections for the main text is not caused by the cancellation of different masses. Figure \ref{figure:localdensity} shows the sample split by local environment. The 2-D local density $\delta_{\rm 2D}$ is estimated by the five nearest neighbor galaxies. Again, we do not see dependence on $\delta_{\rm 2D}$. Finally, in Figure \ref{figure:A_NLA_sys} we show the $A_{\rm NLA}$ constraints for various conditions presented in the previous plots. The main results are shown in red. At $z=1.19$, we see systematically positive $A_{\rm NLA}$ at $\sim2\sigma$ levels, while the signal is weaker at $z=1.47$.
Figure \ref{figure:disperse} is an example of the filamentary structure detected by DisPerSe in the SXDS region at $z=1.47$.

\begin{figure*}
\includegraphics[width=0.95\textwidth]{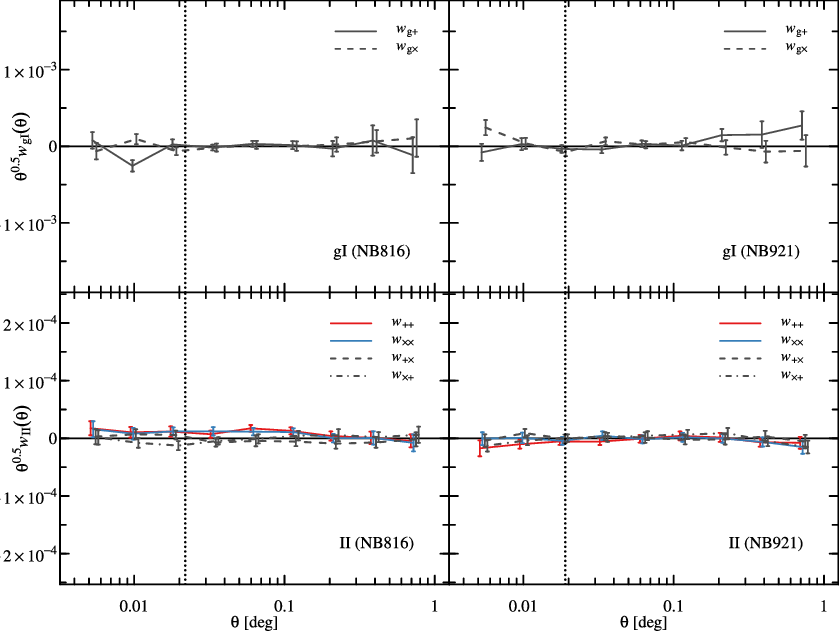}
\caption{The galaxy-PSF cross correlation (top panels) and shape-PSF cross correlation (bottom panels). The $z$-band information is used, and PSF are drawn at the positions of objects. Left panels are for NB816 ($z=1.19$) and right panels are for NB921 ($z=1.47$). The $y$-axis is the same as Figure \ref{figure:IA_all}.
}\label{figure:PSF}
\end{figure*}

Finally, we inspect the presence of a possible contamination from the PSF, by measuring the cross correlations between PSF and galaxies/shapes. Figure \ref{figure:PSF} shows the galaxy-PSF and shape-PSF correlations. Here, PSF are measured at the positions of galaxies. The signal is negligible compared to the galaxy-shape and shape-shape correlations in Figure \ref{figure:IA_all}, indicating the systematic effects from the PSF estimates is negligible.

\vspace{-0.5cm}

\vspace{-0.5cm}

\vspace{0.5cm}
\bibliographystyle{apj.bst}
\bibliography{pasj.bib}

\end{document}